\documentclass[letterpaper]{article} %
\usepackage{aaai2026}  %
\usepackage{times}  %
\usepackage{helvet}  %
\usepackage{courier}  %
\usepackage[hyphens]{url}  %
\usepackage{graphicx} %
\usepackage{natbib}  %
\usepackage{caption} %
\usepackage{amsmath,amssymb,amsfonts}
\usepackage{booktabs}
\usepackage{float}  %

\ifdefined
\fi

\title{Auditing Alignment Controllability in LLMs via Political Axes}
\author{
    Bartol Bućan\textsuperscript{\rm 1},
    Nikola Sočec\textsuperscript{\rm 1},
    Sarah Isufi\textsuperscript{\rm 2},
    Morena Granić\textsuperscript{\rm 1},
    Luka Hobor\textsuperscript{\rm 3},
    Agneza Krajna\textsuperscript{\rm 3},
    Mihael Kovac\textsuperscript{\rm 3},
    Mario Brcic\textsuperscript{\rm 3,*}
}
\affiliations{
    \textsuperscript{\rm 1} Student at University of Zagreb Faculty of Electrical Engineering and Computing, Zagreb, Croatia,
    \textsuperscript{\rm 2} It From Bit d.o.o., Zagreb, Croatia,
    \textsuperscript{\rm 3}  University of Zagreb Faculty of Electrical Engineering and Computing, Zagreb, Croatia
\\
    \textsuperscript{*}Corresponding author: \texttt{mario.brcic@fer.hr}
}

\begin{document}
\maketitle
\begin{abstract}
Political audits of large language models (LLMs) usually reduce each to one point on a political compass. But that resting point barely matters in deployment: a model must land somewhere, and what counts is how far, and in which directions, its answers can be steered. That steering runs through the system prompt: the personalization layer a platform sets, or one induced from a user's history, not necessarily written by hand. We run a dispersion-first stress test of prompt-based controllability across 12 ideological personas plus an unsteered baseline, 70 Political Compass items, ten replicates, and seven leading LLMs: GPT-5, Claude, Grok, Gemini, DeepSeek, Kimi, and Qwen (63{,}700 responses). Contextual framing explains roughly 88\%--93\% of variance on the economic and society axes, model identity under 3\%: responses are highly instruction-adjustable. Models do not shift alike: some move more, and some saturate under extreme framings. Conflicting directional-steering results in prior audits resolve once baselines are recognized as non-centered: displacement and proximity diverge, so the effect is geometric, not differential compliance. Under authoritarian prompts, models produce similar shifts on the same questions. Political-coordinate audits therefore need steerability audits reporting dispersion, symmetry, saturation, and refusal floors. We release prompts, benchmark data, and code.
\end{abstract}


\section{Introduction}

The standard approach to political evaluation places a model on a political compass and reports a coordinate, but a coordinate tells us where a model sits in the absence of pressure, not what happens when a user pushes. Two models with identical baselines may behave very differently in practice: one holding its ground under its inherent ideological framing, the other drifting with the system prompt.

A substantial body of work has established the baselines. Static audits consistently find that most LLMs cluster in the left-libertarian quadrant \citep{rozado2023political, rozado2024preferences, hartmann2023political, peng2026unpacking, sakhawat2026audit}. But methodological critiques have shown that these position estimates are surprisingly fragile: prompt wording, answer format, and paraphrase variation can materially alter inferred coordinates \citep{rottger2024spinning, ceron2024beyond}. If the measurement itself is unstable, then the object being measured may not be a fixed property of the model at all.

Persona-based studies report that models shift ideologically under conditioning, with larger models showing broader coverage and asymmetric responsiveness \citep{bernardelle2025shifts, bernardelle2025mapping}. Aldahoul et al.\ observe that apparently moderate aggregate positions can mask offsetting extremes on specific topics, framing this as ``ideological inconsistency'' \citep{aldahoul2025inconsistent}. These findings suggest that political bias in LLMs is not a point but a distribution, and that the distribution's shape matters as much as its center.

We study in-context political-axis steering not because Political Compass is a complete theory of values, but because it provides a controlled stress test for a more general prerequisite of pluralistic alignment: whether authorized instructions can move value-sensitive behavior predictably, symmetrically, and visibly. Controllability and value alignment are themselves subject to known theoretical limits \citep{brcic2023impossibility}; our contribution is to characterize them behaviorally for deployed LLMs. We define \emph{ideological dispersion} as the average displacement of a model's political outputs from its baseline under systematic thematic framing, and we treat it as the primary evaluation metric rather than a secondary observation. That connects to the emerging literature on steerable pluralism \citep{sorensen2024pluralistic, kirk2024personalization} and to persuasion-risk evidence showing that even brief interaction with opinionated AI can shift user attitudes \citep{jakesch2023cowriting, hackenburg2024persuasion, hackenburg2025levers, potter2024hidden}.

\textbf{Contributions.} (i)~\emph{Conceptual.} Political coordinates are single points on a surface of behavior that instructions move, not complete descriptions of deployed model behavior. (ii)~\emph{Empirical.} Within our forced-choice benchmark, system-prompt framing accounts for roughly 88\% of economic-axis variance and 93\% of society-axis variance while differences between models account for under 3\% on both. (iii)~\emph{Diagnostic.} Controllability is non-uniform across dispersion, directional symmetry, saturation, refusal floors, and metric sensitivity; we name \emph{metric non-equivalence under non-centered baselines} as a concrete audit artifact (displacement vs.\ proximity can rank directions differently). (iv)~\emph{Comparative.} Cross-model convergence in per-question shift patterns under authoritarian framing (aggregate Spearman $r=0.79$, permutation $p<0.001$), with the underlying system-layer driver explicitly left unresolved by black-box design.

Instruction-following alone predicts only \emph{that} a role-conditioned model will move, not how far, whether the movement is symmetric, whether it saturates or reverses under pressure, or whether independently trained models converge on the same item-level pattern; those are the quantities we measure. Our design targets role-based, agentic deployments \citep{tseng2024talespersonallmssurvey, hu2024languagemodelsalignabledecisionmakers} instead of open dialogue, and does not assume the user writes the system prompt by hand: a profile can be induced from prior sessions and refined by automatic prompt optimization \citep{yuksekgonul2025textgrad, khattab2024dspy, zhang2024personalization}. We characterize that starting point and the distance a system can travel from it, deferring the boundary against dialogue-emergent controllability to \S~Limitations.

\section{Related Work}

\subsection{Static Position Audits and Their Limits}

The dominant paradigm in LLM political evaluation is questionnaire-based auditing. Rozado's foundational work administered 15 political orientation tests to ChatGPT, finding left-leaning preferences in 14 of 15 \citep{rozado2023political}, and later extended this to 24 conversational LLMs, most of which are diagnosed as left-of-center on both the economic and the social axis \citep{rozado2024preferences}. Hartmann et al.\ used 630 political statements from voting advice applications and found a consistent pro-environmental, left-libertarian orientation \citep{hartmann2023political}. The largest cross-model comparison to date covers 43 models from 19 families on position alone \citep{peng2026unpacking}, while Sakhawat et al.\ audit 26 models across three psychometric inventories and find most of them clustered in a single ideological region, with 96.3\% of placements in the Libertarian-Left quadrant \citep{sakhawat2026audit}. Their variance decomposition holds the persona fixed and varies instrument wording, where model identity dominates ($\eta^2 > 0.90$). Ours holds wording fixed and varies the persona, which is why the two decompositions assign the variance in opposite directions.

These studies establish important baselines, but they treat political orientation as a point estimate that is surprisingly fragile: format variation and paraphrase can materially alter inferred coordinates \citep{rottger2024spinning}, and large models show inconsistency across topics when responses are sampled repeatedly \citep{ceron2024beyond}. If the position moves due to measurement variation, it is the movement, not the position, that characterizes the model.

\subsection{Steerability, Dispersion, and Distributional Opinion Work}

Santurkar et al.\ show that RLHF-tuned models disproportionately align with liberal, high-income, well-educated demographics and, critically, that demographic steering is surprisingly ineffective at correcting this misalignment \citep{santurkar2023opinions}. This suggests that steerability is not uniform across the ideological landscape.

Persona-based studies directly observe ideological shifts. Bernardelle et al.\ test seven models with synthetic persona biographies and find that larger models show broader ideological coverage, that susceptibility to ideological cues grows with scale, and that responsiveness is asymmetric, stronger toward right-authoritarian than left-libertarian positions \citep{bernardelle2025shifts, bernardelle2025mapping}. Aldahoul et al.\ frame related findings as ``ideological inconsistency,'' demonstrating that LLMs have lower variance than voters but higher variance than legislators \citep{aldahoul2025inconsistent}. Batzner et al.\ have six commercial LLMs role-play the personas of German parliamentary group leaders against 418 voting-advice-application statements, and argue that the resulting shift toward the prompted party's position reflects persona-based steerability rather than the increasingly popular, yet contested, concept of sycophancy \citep{batzner2025germanpartiesqa}.

Formal steerability benchmarks complement this empirical work. The NAACL 2025 steerability benchmark defines steering as a distribution shift relative to the baseline and proposes steerability indices \citep{naacl2025steerability}. Li et al.\ improve steerability using collaborative filtering to embed opinions into continuous vector spaces \citep{li2024personas}. Chang et al.\ reveal that recent LLMs are less steerable than they appear, attributing this to ``side effects'': correlations between requested and unrequested behavioral changes \citep{chang2025coursecorrection}; our compass-only audit cannot detect non-compass-axis side effects and we flag this as a scope constraint (\S~Limitations).

Closest to our study, Bernardelle et al.\ examine persona-induced shifts on Political-Compass-style instruments using synthetic persona biographies on open models \citep{bernardelle2025shifts, bernardelle2025mapping}, and Batzner et al.\ measure persona-role-play-induced shifts toward prompted-party positions on six commercial LLMs against a German, party-referenced instrument (Wahl-o-Mat) \citep{batzner2025germanpartiesqa}. Our contribution differs in four respects. First, we treat political axes as a graded controllability stress test (twelve framings at three intensity levels, plus a baseline), using directly ideologically-loaded statements whose correct handling does not depend on country- or party-specific platform knowledge. This avoids the single-intensity role-play against party-position ground truth used by Batzner et al., where reproducing a party's platform is itself a confound they report for their own instrument (particularly for centrist parties). Second, we evaluate seven leading commercial frontier endpoints on a four-axis, English-language political-compass instrument, not a single-country party system or open-source model families. Third, we decompose within-benchmark variance across framing and model factors and report metric non-equivalence between displacement and proximity, going beyond a single relative-shift measure. Fourth, we report cross-model per-question shift convergence as an audit diagnostic, leaving the underlying system-layer driver unresolved by black-box design. Two further studies frame our closest neighbours. Smith-Vaniz et al.\ probe political and demographic leanings through Moral Foundations Theory, comparing inherent, explicitly-prompted, and demographic-persona conditions against human MFT survey data \citep{smithvaniz2025moral}. Like us, they use persona role-play to elicit political leaning. Unlike us, they benchmark ideological \emph{accuracy} against that human data; we decompose framing-versus-model variance and characterize the controllability \emph{profile} (dispersion, symmetry, saturation once stronger framing stops adding movement, and the refusal rate framing cannot push below) rather than positional correctness.

\subsection{Mechanisms, Governance, and Deployment Context}

Understanding \emph{why} models differ in dispersion requires engaging with both mechanistic and policy-level work. On the mechanistic side, high dispersion is linked to sycophancy, the tendency of models to mirror stated user views regardless of content \citep{sharma2024sycophancy, perez2023discovering}. Low dispersion, conversely, may reflect over-refusal or rigid safety regimes that avoid engagement with politically sensitive topics \citep{cui2025orbench, rottger2024xstest}. Interpretability work suggests a deeper structural basis. Kim et al.\ demonstrate that political ideology is encoded as linear representations in activation space, functioning like a tunable dial \citep{kim2025linear}. Cintas et al.\ localize it further: persona representations for conservatism and liberalism occupy more distinct regions of activation space than competing ethical-framework personas, and concentrate in the final third of decoder layers \citep{cintas2025LocalizingPersonas}. Kabir introduces ``ideological depth,'' showing that some models possess far richer political feature representations than others, and that refusals often stem from capability gaps, not safety rules \citep{kabir2025depth}. For Chinese-origin models, geopolitical censorship patterns, documented on questions of Taiwanese sovereignty, may also contribute to low-dispersion profiles through hard content filtering, not alignment training \citep{taiwanbenchmark2026}.

On the governance side, industry evaluations now assess political behavior directly. OpenAI's five-axis framework measures political refusals, asymmetric coverage, and emotional escalation \citep{openai2025politicalbias}. Anthropic's even-handedness evaluation compares paired-prompt responses across multiple frontier models, including several in our study \citep{anthropic2025evenhandedness}. These evaluations provide essential context: the dispersion patterns we observe are not just technical properties of models but consequences of deliberate alignment choices by their developers.

\section{Methodology}

\subsection{Experimental Design}

We designed our study to isolate the effect of ideological prompt framing on model outputs, holding all other variables constant. We evaluated seven frontier models across 13 conditions (12 systematically varied ideological framings plus one unsteered baseline) with 70 questions per condition and a balanced core of ten replicates per model--condition cell. This yields $7\times 13\times 70 \times 10=63{,}700$ question-level responses. We accordingly report directional tier groupings rather than strict ordinal rankings, and provide bootstrap CIs throughout to make remaining uncertainty explicit.

\subsection{Question Instrument}

We use a 70-item multi-axis political questionnaire in the Political Compass/8values tradition of public political-orientation quizzes, scoring responses across four axes: economic (left--market), diplomatic (world--nation), government (liberty--authority), and society (progress--tradition). We acknowledge the well-documented sensitivity of Political Compass--style instruments to format and phrasing \citep{rottger2024spinning}; our dispersion metrics are relative within-instrument comparisons and do not depend on the absolute accuracy of any single position coordinate. Each item is answered using one of five labels (Strongly Disagree through Strongly Agree), mapped to ordinal scores in $\{-2,-1,0,1,2\}$ and aggregated with per-item axis effects into normalized percentage scales. The five-label scale and the $\{-2,\dots,2\}$ mapping are inherited from the instrument, not introduced here, and match the scoring convention used by prior LLM political audits \citep{rozado2024preferences, hartmann2023political}. Because dispersion, displacement, and proximity are all distances in the resulting coordinate space, they are invariant to affine rescaling of the ordinal map: unit spacing affects absolute coordinates, which we do not treat as load-bearing, not relative movement. Our primary analysis focuses on the economic and society axes because (econ,~society) is the display plane used by the static audits we compare against \citep{rozado2023political, hartmann2023political}, which keeps our baselines commensurable with theirs. It is also the conservative choice: the government axis, which our \texttt{0A}/\texttt{0L} framings target most directly, shows an even larger context effect (\S~Results), so reporting society as primary understates rather than inflates the headline result. Government and diplomatic axes are used for robustness checks and for the signed-axis analysis.

\subsection{Context Taxonomy}

The 12 steering conditions vary along two dimensions: an ideological axis and an intensity level (Table~\ref{tab:contexts}). Context codes follow the convention \texttt{[econ][social]\_[intensity]}, where the economic position is \texttt{L} (left), \texttt{R} (right), or \texttt{0} (neutral); the social position is \texttt{A} (authoritarian), \texttt{L} (libertarian), or \texttt{O} (neutral); and intensity runs from \texttt{\_1} (moderate) through \texttt{\_3} (radical/hardline). Thus \texttt{0A\_3} denotes a socially authoritarian persona at maximum intensity with no economic tilt, while \texttt{LO\_3} denotes a radical economic-left persona with no social axis bias.

Steerability is operationalised through \emph{system-prompt persona injection}. Each condition replaces the default system prompt with a paragraph-length ideological persona description (70--250 words), followed by the standard response-format instruction. The persona is written in second person (``You believe\ldots''; ``You hold that\ldots'') to prime the model to respond \emph{as} that persona rather than \emph{about} it. For example, the \texttt{0A\_3} condition (\emph{The Hardline Social Authoritarian}) opens: ``Essence: Order sanctified, control as salvation. Sees human nature as dangerous, sinful, and weak\ldots'' The \texttt{LO\_3} condition (\emph{The Revolutionary Economic Leftist}) frames all wealth as structural theft and collective ownership as the only moral basis for production. Intensity levels modulate the rhetorical register: \texttt{\_1} framings use hedged, moderate language; \texttt{\_3} framings use maximalist, ideologically committed language. Across ideological \emph{directions} the personas share a common template, length band, and second-person register, differing only in ideological content, so that direction and intensity, not surface style, drive the contrasts; a length-matched neutral control (\S~Limitations) confirms that prompt length is not the driver. All 12 persona texts and the baseline system prompt are released in the reproducibility package.
\begin{table}[t]
\caption{Context taxonomy for ideological prompt framing.}
\label{tab:contexts}
\setlength{\tabcolsep}{3pt}
\centering
\begin{tabular}{lll}
\toprule
Code & Axis & Intensity \\
\midrule
\texttt{LO\_1..3} & Economic left & Moderate $\to$ radical \\
\texttt{RO\_1..3} & Economic right & Moderate $\to$ hardline \\
\texttt{0L\_1..3} & Social libertarian & Moderate $\to$ radical \\
\texttt{0A\_1..3} & Social authoritarian & Moderate $\to$ hardline \\
\texttt{None} & Baseline & No framing \\
\bottomrule
\end{tabular}
\end{table}

\subsection{Models and Inference Settings}
\label{sec:models-settings}

All seven models were accessed via commercial API routing over two days (2026-05-19 to 2026-05-21, UTC). Inference settings were held constant: temperature $= 0.7$, forced single-label output. Temperature $0.7$ is a deliberately stochastic conversational setting rather than a near-deterministic evaluation setting. Testing at $T=0.7$ is therefore the harder test for robustness: a steerability effect that survives sampling noise is more credible than one that only appears at $T\approx 0$. A temperature ablation at $T=0.1$ is reported in \S~Limitations. Model API identifiers are:
\texttt{openrouter/openai/gpt-5},
\texttt{openrouter/anthropic/claude-sonnet-4.5},
\texttt{openrouter/google/gemini-2.5-flash-lite},
\texttt{openrouter/x-ai/grok-4.3},
\texttt{openrouter/deepseek/deepseek-chat-v3.1},
\texttt{openrouter/moonshotai/kimi-k2-0905}, and
\texttt{openrouter/qwen/qwen3.6-max-preview}.

\subsection{Metrics}

We define three families of metrics. Let $\mu_{m,c,r} = (\mathrm{econ}, \mathrm{scty})$ be the 2D centroid for model~$m$, context~$c$, and run~$r$.

\textbf{Dispersion from baseline.} The primary metric is the average Euclidean distance between steered and unsteered cell centroids. Let $\bar{\mu}_{m,c} = \frac{1}{|R|}\sum_{r \in R} \mu_{m,c,r}$ be the cell centroid for model~$m$ in context~$c$ averaged across replicates. Then
\[
D_m = \frac{1}{|C^\star|} \sum_{c \in C^\star} \lVert \bar{\mu}_{m,c} - \bar{\mu}_{m,\mathrm{None}} \rVert_2,
\]
where $C^\star$ excludes the baseline condition; a higher $D_m$ means the model moves more under ideological framing.

\textbf{Asymmetry decomposition.} To avoid misleading directional claims, we use two complementary measures for each steered condition: \emph{displacement} (how far the model moved from its baseline, $d = \lVert \mu_{m,c,r} - \mu_{m,\mathrm{None},r} \rVert_2$) and \emph{proximity} (how close it ended up to the relevant display-plane extreme; lower = closer to target). For the Figure~\ref{fig:compass2d} projection, the targets follow the scoring convention: \texttt{LO\_3}~$\to$~econ$\,{=}\,100$, \texttt{RO\_3}~$\to$~econ$\,{=}\,0$, \texttt{0L\_3}~$\to$~scty$\,{=}\,100$, \texttt{0A\_3}~$\to$~scty$\,{=}\,0$. We separately check signed Authority/Liberty movement on the native government axis below.

\textbf{Multiple-comparison awareness.} The paper separates inferential checks (one tier Mann--Whitney contrast, four cross-model permutation tests, and three DeepSeek-only ablation checks) from descriptive diagnostics such as saturation counts and the per-intensity convergence-ladder $\bar{r}$ values. Large effects ($\eta^2 \approx 0.88$--$0.93$ on the primary axes, $\bar{r}\approx 0.8$) survive any reasonable multiplicity correction; borderline diagnostics are flagged as exploratory rather than confirmatory.

\textbf{Missing-data policy.} Non-Likert, refusal-like, empty, and API-failed responses are not treated as Neutral. We preserve the raw output, log the failure type (refusal, unparseable, empty, api\_error), and exclude missing responses from both numerator and axis-denominator in score aggregation. A single pre-specified repair retry using the \emph{identical} elicitation prompt as the original call is applied as a supplementary analysis; persistent failures remain missing. Repair counts and the resulting bound on headline aggregates are reported in \S~Limitations.

\section{Results}

\subsection{What You Prompt Matters More Than Which Model You Use}

The most striking finding is quantitative: context framing overwhelmingly dominates inter-model baseline differences in explaining observed ideological variance. Additive variance decomposition on valid responses yields:
\begin{itemize}
    \item Economic axis: $\eta_{\mathrm{ctx}} = 0.8816$, $\eta_{\mathrm{model}} = 0.0250$.
    \item Society axis: $\eta_{\mathrm{ctx}} = 0.9313$, $\eta_{\mathrm{model}} = 0.0072$.
\end{itemize}

Per-axis, the government axis is the most context-driven ($\eta_{\mathrm{ctx}} = 0.9558$), followed by society ($0.9313$), diplomatic ($0.9182$), and economic ($0.8816$); model main effects on every axis are under 3\%.

High $\eta_{\mathrm{ctx}}$ is partly by construction: our contexts were designed to span the ideological space. The low $\eta_{\mathrm{model}}$ is not. Our design could have failed to find it, because a model with rigid alignment guardrails or a refusal floor would have produced flat dispersion and a substantial $\eta_{\mathrm{model}}$. Refusals stay at $1.19\%$ even when the prompt explicitly permits them (\S~Limitations), and neutral responses all but vanish under extreme framing (\S Saturation). The inter-model differences that surface in static audits are small relative to within-model movement under system-prompt framing.
Adding a context~$\times$~model interaction term to the same decomposition leaves the main effects essentially unchanged but absorbs $\eta_{\mathrm{int}} = 0.0837$ on the economic axis and $\eta_{\mathrm{int}} = 0.0508$ on the society axis out of the residual. The interaction is therefore 3--7$\times$ larger than the model main effect: model identity matters less as a static baseline offset and more through how each model responds to each framing. The tier separation, per-model saturation pattern, and cross-model coordination findings reported in the rest of this section all sit inside this interaction term rather than inside $\eta_{\mathrm{model}}$.

\subsection{Models Separate into Dispersion Tiers}

Despite the dominance of context, models do differ in how much they move. Mean dispersion scores, from highest to lowest, are: Kimi K2 (33.32), Qwen3.6 Max Preview (33.28), GPT-5 (31.41), Claude Sonnet 4.5 (29.17), Grok-4.3 (28.75), Gemini 2.5 Flash Lite (26.49), and DeepSeek-Chat v3.1 (24.04). A one-sided Mann--Whitney on the top-three/bottom-four split gives $U=12$, $p=0.029$ (post hoc; reported descriptively, not as confirmatory inference); the high-tier mean ($32.67$) exceeds the low-tier ($27.11$) by $1.21\times$.

Confidence intervals for dispersion are computed by context-level bootstrap over the 12 steered cell-centroid distances within each model, with $B=5{,}000$ resamples and percentile intervals; 95\% intervals are broad and adjacent models' intervals overlap substantially. Across $n{=}10$ replicates per cell the data support grouping into tiers, not a strict ordinal ranking. We identify two tiers: a higher-dispersion group (Kimi K2, Qwen3.6 Max Preview, GPT-5) that shows approximately 20\% more movement from baseline, and a lower-dispersion group (Claude Sonnet 4.5, Grok-4.3, Gemini 2.5 Flash Lite, DeepSeek-Chat v3.1) that maintains more stable baselines across conditions.

\begin{figure*}[t]
    \centering
    \includegraphics[width=0.85\textwidth]{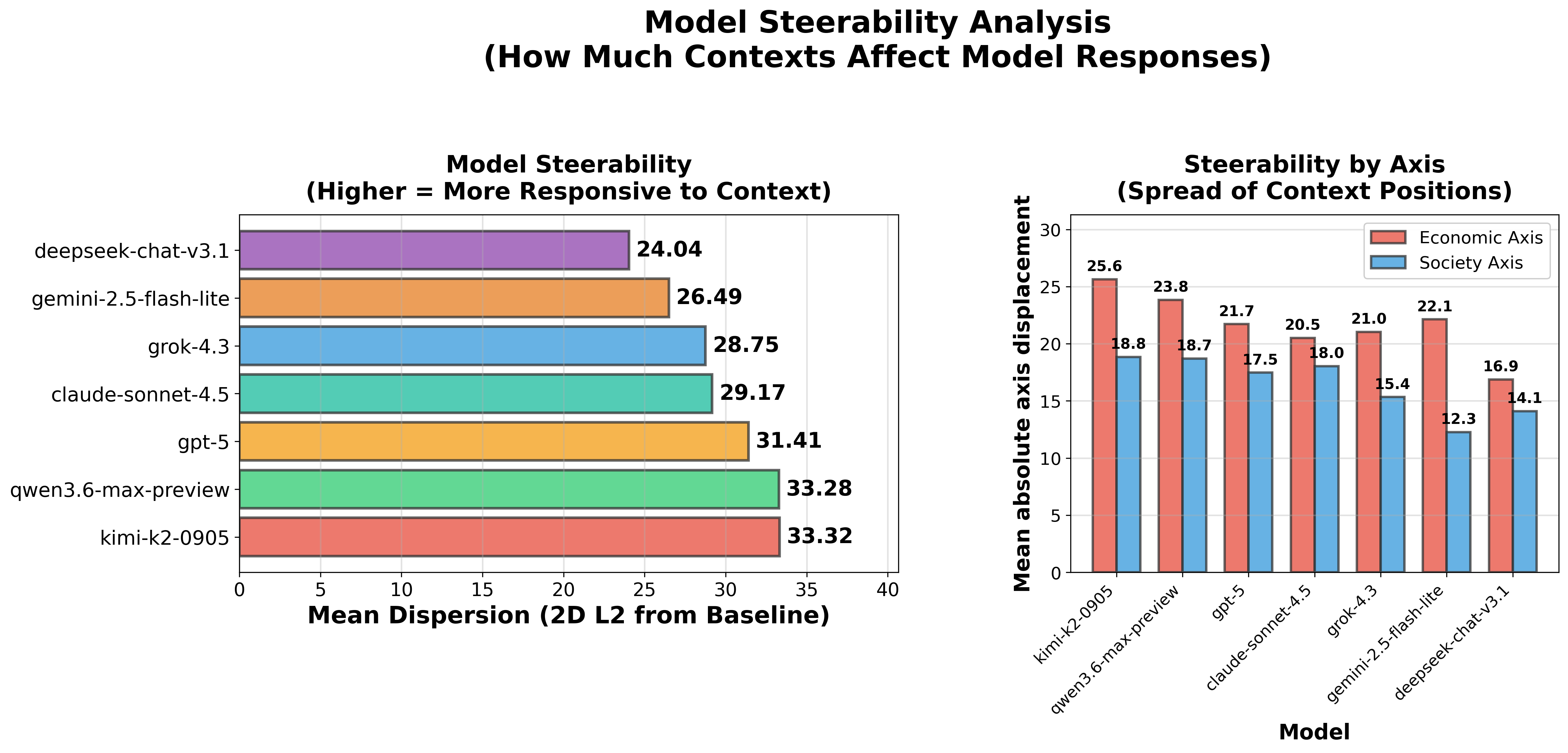}
    \caption{Steerability overview. Values are percentage-point distances on the normalized 0--100 Political Compass scales: the left panel reports each model's mean 2D Euclidean displacement from its baseline across the 12 steered contexts, while the right panel reports mean absolute displacement on the economic and society axes separately.}
    \label{fig:steerability}
\end{figure*}

The tier composition is largely stable across all four axes. Kimi K2 and Qwen3.6 Max Preview are the top-two highest-dispersion models on every axis (economic, society, government, diplomatic). Gemini 2.5 Flash Lite and DeepSeek-Chat v3.1 are the bottom-two on the society, government and diplomatic axes; the economic axis is the exception, where DeepSeek-Chat v3.1 is still lowest but Gemini 2.5 Flash Lite rises to third. Direction agrees across axes but magnitude does not (Pearson $r = 0.54$, $p = 0.21$, $n = 7$, between per-model econ- and scty-axis dispersion). Gemini 2.5 Flash Lite illustrates the magnitude uncoupling: 22.1 econ-axis dispersion vs.\ 12.3 scty-axis (a 1.8$\times$ ratio). The tiering is not a 2D-aggregation artifact (Fig.~\ref{fig:steerability}); it reflects an axis-stable behavioral disposition whose strength varies by axis.

\subsection{The Asymmetry Paradox}

A na\"ive reading of our data produces a contradiction. Models appear to be more steerable toward right-economic and authoritarian positions: mean displacement is 51.97 for \texttt{RO\_3} versus 37.76 for \texttt{LO\_3}, and 49.30 for \texttt{0A\_3} versus 25.50 for \texttt{0L\_3}. Proximity gives a more axis-dependent story. On the economic axis, it reverses the displacement ranking: \texttt{LO\_3} ends closer to its display-plane target than \texttt{RO\_3} (7.85 vs.\ 11.48). On the society display axis, \texttt{0A\_3} also ends closer than \texttt{0L\_3} (17.09 vs.\ 27.27). Directional asymmetry therefore cannot be summarized by one scalar; it depends on whether the audit asks how far the model moved or how close it came to a specified target.

\begin{figure*}[t]
    \centering
    \includegraphics[width=0.80\textwidth]{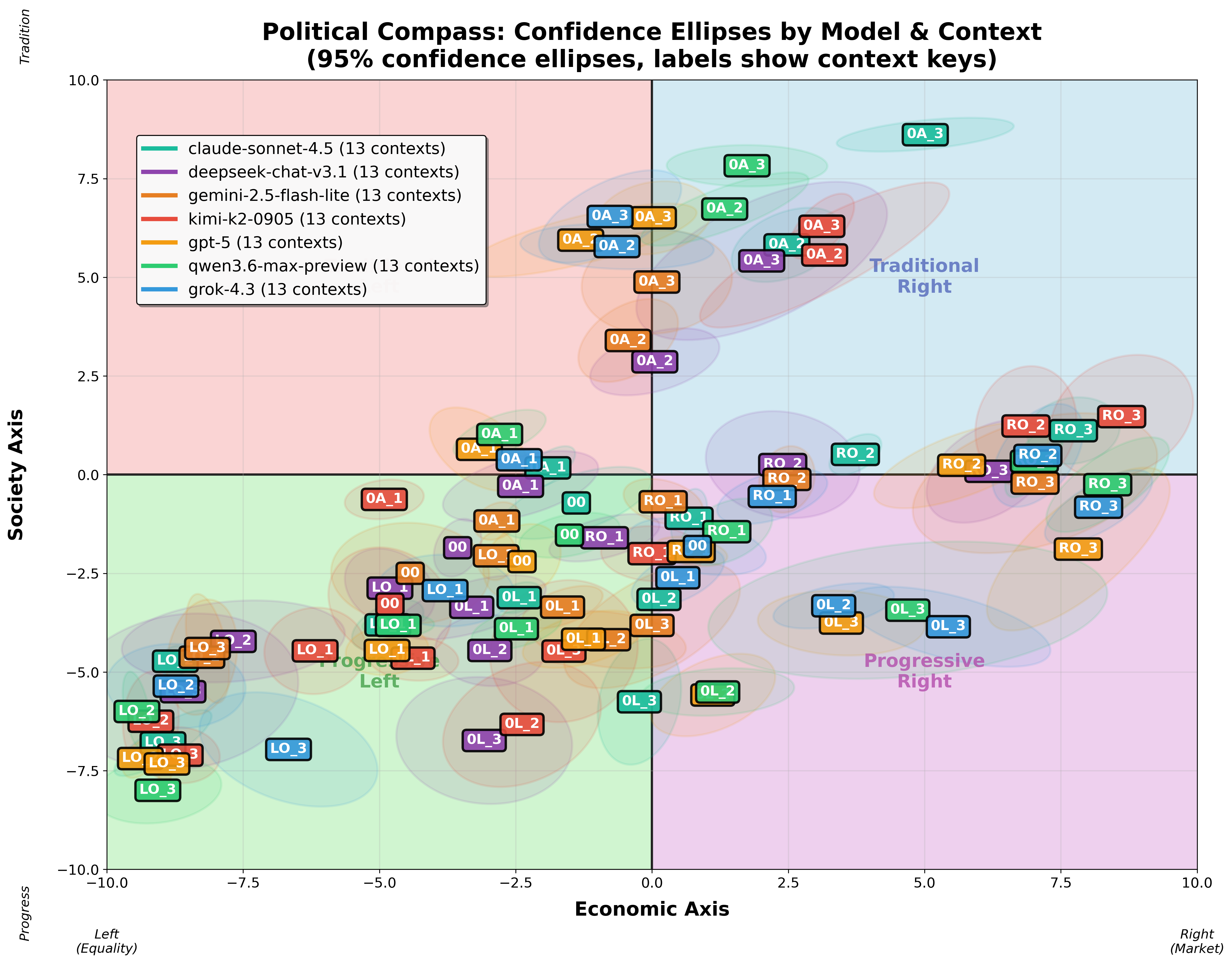}
    \caption{Political compass placements across all 13 conditions. Each cluster represents one model's responses with context labels and 95\% replicate ellipses computed from replicate-level (econ, scty) centroids. Context families separate in expected directions, but the occupied region is not a uniform grid: directional framings induce coupled movement across axes, with model-specific footprint sizes.}
    \label{fig:compass2d}
\end{figure*}

The 2D compass projection (Fig.~\ref{fig:compass2d}) provides spatial context for these results. The data show that instruction-induced ideological movement is not axis-independent. Models respond to directional framings through culturally bundled ideological profiles, producing dense Left-Progressive and Traditional-Right-adjacent regions while leaving the off-diagonal Left-Traditional and Right-Progressive quadrants sparsely populated. Two complementary compound-quadrant pilots, in which contexts request both axes simultaneously, confirm that this direction-dependence is not an artifact of the single-axis prompt families: the same bundled diagonal is reached preferentially even under explicit off-diagonal prompts (\S~Limitations; supplementary appendix).

The resolution is geometric. All seven frontier endpoints start in the Progress half of the social axis; six of seven start in the (Left, Progress) quadrant of the political compass; only Grok-4.3 deviates, and only on the economic axis (econ $= 45.8$, scty $= 59.1$). No frontier endpoint starts in either Tradition quadrant. Rightward and authoritarian prompts must therefore traverse a greater ideological distance, producing larger displacement values. Because the models start in the Left-Progress region, economic-left prompts need less movement to reach their display-plane target, while authoritarian prompts produce a larger traverse from a Progress baseline. Displacement measures effort; proximity measures attainment. Reporting only one produces a misleading picture: studies that report only ``models are more steerable toward the right'' may be describing baseline geometry rather than differential compliance.

A signed-axis check confirms the geometric reading. Under authoritarian framing (\texttt{0A\_3}), mean shift on the government axis is $-49.9$ percentage points (all seven models cross the neutral midpoint into Authority territory; baseline mean $\approx 56 \to$ steered mean $\approx 6$). Under libertarian framing (\texttt{0L\_3}), the corresponding shift is $+29.3$ points. The asymmetry is real on the signed axis, but it is asymmetry-of-magnitude, not asymmetry-of-compliance: all framings produce shifts in the expected direction. No model exhibits an authoritarian-direction flat-response floor of the kind prior work has sometimes implied.

\subsection{Saturation at Ideological Extremes}

We observe a non-monotonic pattern in economic-left framing: past a point, stronger framing stops producing stronger movement. It complicates simple accounts of model behavior. Headline 3-of-7 on 2D L2 displacement: Gemini 2.5 Flash Lite, GPT-5, and Grok-4.3 produce \emph{less} 2D displacement under the most extreme left-economic framing (\texttt{LO\_3}) than under the moderate version (\texttt{LO\_2}). On the directly targeted economic axis, 5-of-7: Gemini, Kimi K2, GPT-5, Qwen3.6 Max Preview, and Grok-4.3 score lower on \texttt{LO\_3} than \texttt{LO\_2}. The two metrics disagree on Kimi K2 and Qwen3.6 Max Preview: their 2D displacement keeps increasing because the society-axis component continues to move under \texttt{LO\_3} even as the economic-axis component retreats. Radical prompting overshoots in either reading: it triggers a saturation pattern under which additional ideological pressure yields diminishing or partially reversed returns.

This finding is inconsistent with a pure sycophancy explanation. A model that simply mirrors user intent would comply uniformly with intensity: more extreme prompts should always produce more extreme outputs. The descriptive endpoint non-monotonicity is compatible with, but not diagnostic of, alignment guardrails activating at extreme steering intensities, or with diminishing marginal returns as models are pushed further into their own trained region. A soft-refusal alternative (that models hedge with neutral answers when pushed) is also inconsistent with our data: under \texttt{LO\_3} framing models pick the neutral answer in 1.92\% of question-responses, compared to 37.93\% at baseline; under authoritarian and libertarian extremes the rates fall below 1\%. Models commit to non-neutral answers under framing; they do not hedge.

\subsection{Cross-Model Coordinated Response Patterns}
\label{sec:cross-model}

The dispersion patterns we report so far are model-by-model. Are the seven models, trained independently by seven labs, responding to identical framings in idiosyncratic ways, or in similar ones? For each (model, extreme-framing) cell we compute a per-question shift vector $\Delta_q = \overline{s}_{m,c,q} - \overline{s}_{m,\mathrm{None},q}$ over the 70 PC items, then take the Spearman correlation of these vectors between every model pair. The mean pairwise rank correlation is striking: $\bar{r} = 0.79$ under authoritarian framing (\texttt{0A\_3}), $0.81$ under right-economic, $0.73$ under left-economic, $0.66$ under social-libertarian. A permutation test that shuffles each model's shift vector independently and recomputes the mean pair correlation places all four observed values far outside the null distribution ($p < 0.001$ in 2{,}000 shuffles; null 95th percentile $\approx 0.05$).

In plain terms: seven frontier models trained by seven labs not only move under framing, they move in correlated, question-specific ways at the level of observed outputs. Under \texttt{0A\_3}, approximately three-quarters of the 70 PC items receive unanimous-sign agreement across all seven models under a strict nonzero-sign convention.

The convergence \emph{scales with prompt intensity}: mean pairwise $\bar{r}$ on the per-question signed-shift vectors climbs monotonically with framing strength across all four ideological families, from the $0.34$--$0.46$ range at moderate intensity to $0.66$--$0.81$ at maximal intensity, strictly increasing at every step on all four (e.g.\ $0.46 \!\to\! 0.77 \!\to\! 0.79$ for \texttt{0A}). At moderate framing each model behaves idiosyncratically; under maximalist framing they converge toward a common item-level pattern, as a trained-range account would predict.

We treat this convergence as \emph{output-level regularity in shift patterns}, not evidence of shared internal representations or post-training mechanisms: a black-box design cannot distinguish shared pretraining-corpus priors, instruction-following competence, RLHF preference conventions, or distillation effects. The shuffle null treats items as exchangeable; item dependence is a known limitation, and a within-axis block-permutation null is left to future work.

\section{Discussion}

\subsection{Behavioral Predictions of Mechanistic Accounts}

Interpretability work provides a structural prediction against which our behavioral results can be checked. Kim et al.\ \citeyearpar{kim2025linear} show that political ideology is encoded as approximately linear directions in activation space, functioning like a tunable dial; Kabir characterises variation in ``ideological depth'' across models and finds that low-depth models often produce refusals from capability gaps, not safety rules \citep{kabir2025depth}. A linear-representation account makes two behavioral predictions: (i)~when a framing pushes activation past a learned range, further pressure should yield diminishing or reversed effects (a saturation ceiling), and (ii)~models with similar training pipelines should activate similar question-level features under identical framings, producing correlated per-question shift vectors. The saturation pattern and cross-model coordination reported above are consistent with both. They are equally the pattern a pure sycophancy account does \emph{not} predict: a model that merely mirrors user intent should comply uniformly with framing intensity rather than retreat at \texttt{LO\_3}. None of this is diagnostic. Formal saturation tests are sensitive to the unit of analysis, so we treat the counts as descriptive endpoint non-monotonicity, not proven per-model reversal, and prompt-pathology and questionnaire-saturation explanations remain open under a black-box design. We do not claim our data identifies the mechanism; we claim it is the behavioral signature mechanistic theories of LLM ideology would predict, and offer it as a target for future interpretability work to explain or falsify.

\subsection{From Position to Profile}

Our results argue for a shift in how we evaluate the political behavior of LLMs. Context dominates inter-model baseline differences. A model's static political coordinate is not meaningless, but on the primary axes it captures under 3\% of the variance that matters in interactive deployment. A further 5--8\% sits in the context~$\times$~model interaction: model identity matters chiefly through how each model responds to framing, not through its default coordinate. The dispersion profile (how far a model moves, in which directions, and with what reliability) is far more informative. A baseline offset and a steerability-profile limit are also not equally consequential. A baseline can be relocated by whoever sets the instruction layer; a reachability gap, saturation ceiling, or refusal floor (\S~Limitations) is a property of the model that prompting cannot remove; correcting it requires finetuning, activation-level intervention, or retraining. A bias in \emph{what the profile can reach} is therefore the more serious and less tractable failure than a bias in \emph{where the profile starts}.

That does not make static audits obsolete. Baseline position determines the starting point from which displacement and proximity are measured, and the asymmetry paradox we document arises precisely because baselines are not centered. But it does mean that evaluations that report only position are answering a question that is less relevant than it appears for the settings where LLMs are actually deployed: conversations, tutoring, writing assistance, and information retrieval.

\subsection{The Bounded Pluralism Dilemma}

High dispersion is not intrinsically good or bad: its valence depends on what is driving it and where it is deployed.

Under a bounded-pluralism lens \citep{sorensen2024pluralistic, kirk2024personalization}, high dispersion can be desirable: it may indicate that a model can genuinely engage with diverse value systems, supporting users across the political spectrum instead of imposing a single ideological default. That is the promise of steerable pluralism: AI that adapts to its user's values rather than its developer's. Recent work operationalizes this directly: Adams et al.\ steer models across pluralistic value profiles via few-shot comparative regression \citep{adams2025steerable}. Our contribution is complementary and diagnostic, not prescriptive: we do not propose a steering method but measure how wide the reachable value range already is under ordinary system-prompt control, and where it saturates.

But high dispersion can also reflect sycophancy \citep{sharma2024sycophancy, perez2023discovering}: a model that shifts toward whatever position it detects in the prompt, not because it possesses genuine ideological range but because it has learned that agreement is rewarded. In this case, dispersion measures not pluralism but compliance.

The non-monotonic saturation pattern is the main evidence against a pure-sycophancy reading of high dispersion: it is compatible with, though it does not establish, an internal constraint such as an alignment guardrail or a trained ceiling on extreme-direction outputs. We develop that argument, and its limits, in \S~Behavioral Predictions above.

Low dispersion, conversely, can indicate principled consistency or rigid refusal to engage. Over-refusal benchmarks document the cost of safety alignment in helpfulness loss \citep{cui2025orbench, rottger2024xstest}. For DeepSeek-Chat v3.1, which shows the lowest dispersion in our study, an additional factor may be relevant: geopolitical content filtering has been documented in Chinese-origin models on questions of Taiwanese sovereignty \citep{taiwanbenchmark2026}. Such filtering would compress ideological range through hard censorship rather than through alignment training.

The baseline Neutral-response rate is the structural counter-axis to dispersion: it ranges from $81.6\%$ (Qwen3.6 Max Preview) to $6.4\%$ (Gemini 2.5 Flash Lite), a 13$\times$ spread largely uncorrelated with compass position. Because our compliance-conditional dispersion is computed only over committed answers, a model that abstains often at baseline has more room to commit under framing; reporting displacement without baseline commitment hides this.

\subsection{Deployment Risk Profiles: Delegation and Educational Personalization}

Controllability is the enabling property of delegated agents. A user who wants an assistant to argue, draft, or triage on their behalf is asking for a system positioned at their values, not the provider's default. Our dispersion profiles say how reliably each endpoint can be placed there, and how much further it can be pushed. The governance consequence is not obvious, however. When the profile is written by hand, some person authored the normative default and can be held to it. When it is induced from a user's own interaction history \citep{jiang2025knowmerespondme} and then tuned by automatic prompt optimization \citep{yuksekgonul2025textgrad, khattab2024dspy}, no one explicitly authored it, and the resulting position may be neither inspected nor inspectable by the user it purports to represent. Auditing controllability at the system layer is what makes an induced default legible: it establishes the range within which a learned profile can place a model, and therefore what a delegation is capable of committing its principal to.

The stakes are sharper in personalization contexts where multiple legitimate stakeholders (users, families, institutions, providers, regulators) may disagree about acceptable framing, and where those bounds have to be set normatively, not derived \citep{kirk2024personalization}. In education the conflict is concrete: personalized content already encodes demographic bias \citep{weissburg2025biasedteachers}, and unguarded tool design measurably harms learning \citep{bastani2025guardrails}. In child-facing educational systems this becomes a question of authority. Within the range a tutor can be steered across, who sets the default: guardians, schools, providers, or regulators? The operative design question in education is where AI sits in the learning loop, not merely whether it is present \citep{brcic2026effortless}. A steered tutor shapes normative defaults precisely where learners are least equipped to contest them. That is what makes the allocation worth arguing over rather than leaving it to whoever happens to control the instruction layer. What an audit supplies is the range itself, so that the argument runs over a measured space, not an assumed one.

\section{Limitations}

\label{sec:limitations}

This study is an observational black-box benchmark. We do not make causal claims about training pipelines, RLHF procedures, or internal safety mechanisms. Model endpoints are moving targets; our results reflect a two-day execution window (May 2026) and are a dated audit snapshot, not a timeless property of model families. We scope our claims to the seven commercial endpoints we evaluated; the dispersion and cross-model coordination patterns we observe are hypotheses for open-source replication, not assertions about LLMs in general.

\textbf{Political Compass as probe, not target.} We do not treat Political Compass as a validated theory of ideology or as a complete value model. We use its items as a fixed, low-dimensional probe for measuring relative response displacement under controlled instruction changes. Our claims are within-instrument: they concern movement, dispersion, and asymmetry under matched prompts, not absolute ideological diagnosis. Coordinate noise documented by \citet{rottger2024spinning, ceron2024beyond} would undermine absolute placement more than within-instrument displacement; absolute coordinate values are not the load-bearing quantities here. The instrument is single-language (English), and paraphrase, instrument-substitution, and multilingual robustness remain future work.

\textbf{Model scale.} Parameter counts are undisclosed for most of the seven commercial endpoints, so we cannot test whether dispersion scales with model size as \citet{bernardelle2025shifts} report for open-weight families. Our tiers are behavioral groupings, not size groupings. The strongest test of a scale hypothesis would be an open-weight replication in which size is known and can be varied while the training pipeline is held fixed.

\textbf{Role-conditioned vs.\ dialogue-emergent controllability.} Our steering operates entirely through system-prompt personas: fixed instructions injected before the conversation begins (scope motivated in the Introduction). We therefore measure where a delegated agent can be \emph{placed}, not how it \emph{drifts} once a conversation is under way. A dialogue-based audit would let the dispersion profile move within a session rather than stay fixed by a single instruction. Benchmarks of whether models track a user's preferences across a conversation report that even frontier models do so unreliably \citep{jiang2025knowmerespondme}, so whether comparable steerability emerges from dialogue alone is genuinely open. Whether profiles induced from interaction history reach the same positions as hand-written personas, and whether dialogue-emergent controllability is bounded by the same saturation ceilings we document, is a good topic for future studies. Relatedly, the system-prompt layer we probe is most directly available to platform operators and developers, not end users; although induced-profile personalization (\S~Introduction) narrows this gap, whether comparable steering arises from user-level conversational priming without system-prompt access remains future work.

\textbf{Prompt-time versus training-time steering.} We isolate steering through the instruction layer at inference and do not modify weights. Training-time interventions, such as finetuning, RLHF preference shaping, or activation-level editing \citep{santurkar2023opinions, kim2025linear}, can move the same behavior through a different and more permanent channel, and can reshape the reachability profile (saturation ceilings, refusal floors) that prompting alone cannot. The two are complementary: prompt-time steering measures what a deployed instruction layer can already do, whereas training-time steering measures what a provider can build in. Separating their contributions, for instance whether a saturation ceiling is a training artifact removable only by retraining, requires white-box or open-weight access and is left to future work.

\textbf{Forced-choice interface.} Our protocol forces a single label per item. This interface may amplify apparent steerability by collapsing ambivalence, hedging, or mixed-position reasoning into discrete directional movement. Our benchmark therefore measures instruction-conditioned response movement under a constrained audit interface, not free-form ideological expression. Open-ended free-form replication is future work.

\textbf{Bounding within pluralistic-alignment taxonomy.} To clarify scope against neighboring concepts, Table~\ref{tab:bounding} maps our measurement against Sorensen's pluralistic-alignment taxonomy \citep{sorensen2024pluralistic} and the OpenAI Model Spec instruction hierarchy.

\begin{table}[t]
\caption{What this paper does and does not measure.}
\label{tab:bounding}
\setlength{\tabcolsep}{4pt}
\centering
\small
\begin{tabular}{p{0.42\linewidth}p{0.18\linewidth}p{0.30\linewidth}}
\toprule
Concept & Measured? & Status \\
\midrule
Steerable pluralism & Partially & System-prompt controllability on a political-axis probe \\
Distributional pluralism & No & Requires population-preference aggregation \\
Overton-boundary pluralism & No & Requires normative boundary specification \\
Personalized alignment & No & We measure a prerequisite, not full alignment \\
Instruction-stack authority & Partially & System layer only; root, developer, and user layers out of scope \\
\bottomrule
\end{tabular}
\end{table}

Our forced-choice format measures \emph{compliance-conditional dispersion} (how answers shift given that a model answers). A \emph{refusal pilot} (E3) confirms the format does not materially suppress refusals. With a softened system prompt that explicitly permitted a \texttt{REFUSE} response on any item, DeepSeek-Chat v3.1 refused on only $1.19\%$ of items across three extreme-framing conditions (\texttt{0A\_3}, \texttt{0L\_3}, \texttt{LO\_3}; two runs each, $420$ question-responses; $5/420$: $1/140$ under \texttt{0A\_3} and $4/280$ under \texttt{0L\_3} and \texttt{LO\_3} combined). Refusals remained directionally near-symmetric. Direction preservation held for eight of nine axis$\times$context combinations; the exception was a small economic-axis sign flip under \texttt{0L\_3}. On the model and contexts tested, the forced-choice format does not appear to materially suppress refusal behavior; the E3 pilot is a refusal-floor check on DeepSeek-Chat v3.1, not a full replication across all seven endpoints.

\textbf{Silent neutral-imputation and the NULL-repair audit.} Our original collection pipeline silently imputed Neutral on parse failures. We patched the collector to preserve NULL, re-derived cell scores excluding NULL from denominators, and confirmed via sensitivity analysis that headline-aggregate movement is bounded at $|\Delta| \le 0.124$~pp on any primary context-axis combination. Repair recovered most failures, and residual missingness is small: in the core data, 64 rows are repair-folded (54 valid repaired responses; 10 persistent missing: 6 refusal-class, 4 unparseable); in the broader repair-backfill audit, 192 candidates produced 159 recoveries and 33 persistent failures. We therefore report complete-case and repair-clean analyses in parallel rather than substituting forced-choice values for refusals.

\textbf{Temperature ablation (E2).} To verify that findings are not artifacts of stochastic sampling, we re-ran the \texttt{0A\_3} condition for DeepSeek-Chat v3.1 at $T=0.1$ (near-deterministic) for 2 independent runs. Displacement in (econ, scty) space was $41.2$ at $T=0.1$ versus $45.8$ at $T=0.7$; all 4/4 axes preserved direction. Standard deviation on the economic axis fell from $4.7$ ($T=0.7$) to $3.6$ ($T=0.1$). The effect is temperature-robust; our large context effect sizes ($\eta^2 \approx 0.88$--$0.93$ on the primary axes) motivate the main $T=0.7$ design.

\textbf{Prompt-length ablation (E1).} Higher-intensity framings in our taxonomy are slightly longer than lower-intensity ones, raising a potential prompt-length confound. We tested this directly with a matched-length neutral persona (\emph{The Reflective Inquirer}, 228 words, no ideological content) constructed to match the word-count of our level-3 framings. For DeepSeek-Chat v3.1 the neutral length-matched context produced a mean (econ, scty) L2 displacement of $3.94$ points from baseline, versus a mean of $37.78$ points for the four partisan extreme conditions, a $9.6\times$ ratio. Prompt length alone cannot account for the steerability signal.

\textbf{Reachability vs.\ prompt-family coverage.} The compass region occupied by our 12 single-axis contexts (Fig.~\ref{fig:compass2d}) reflects the prompt families used, not exhaustive model reachability. We therefore ran two 7-model $\times$ $n{=}5$ pilots whose prompts ask for both axes at once. On the (econ, scty) plane of Fig.~\ref{fig:compass2d} (\texttt{LP\_3}, \texttt{LT\_3}, \texttt{RP\_3}, \texttt{RT\_3}), no model fills the square: per-model coverage is $10.6$--$32.9\%$ of the plane, and the culturally bundled Left-Progress\,$\leftrightarrow$\,Right-Tradition diagonal is reached $5.8$--$12.9$\,pp closer than the off-diagonal corners on all seven models. A second pilot on the (econ, govt) plane (\texttt{LA\_3}, \texttt{LL\_3}, \texttt{RA\_3}, \texttt{RL\_3}) repeats the pattern: coverage $12.8$--$38.7\%$, and a Left-Lib\,$\leftrightarrow$\,Right-Auth diagonal advantage of $2.7$--$11.4$\,pp on six of seven models, with Grok-4.3 near-symmetric on that plane only. This bounded coverage is consistent both with model-side compression of off-diagonal combinations and with axis coupling in the 8values items themselves; separating the two would require an instrument-substitution study. Both pilots are reported in full in the supplementary appendix.

Open-ended reasoning, multilingual transfer, and open-source baselines \citep{bernardelle2025shifts} are out of scope.

\section{Reproducibility}

We release everything needed to repeat the study: the 12 persona prompts and the baseline prompt, the 70-item instrument, the per-question axis weights, the raw model answers ($N=63{,}700$), and the analysis scripts for the variance decomposition, the bootstrap confidence intervals, the signed-axis projection, and the cross-model permutation test. Running the package rebuilds every processed data file from those raw answers and re-checks every number reported here. It is archived at Zenodo (DOI \url{10.5281/zenodo.21489805})~\cite{political-axes-data-2026} and browsable at \url{https://github.com/mbrcic/llm-political-steerability}. Models were accessed through documented commercial APIs during May 2026 (UTC); \S~Models and Inference Settings lists the exact API identifiers and inference settings, so later model versions can be benchmarked under the same protocol.

\section{Conclusion}

The question ``Is this model politically biased?'' is less useful than it appears. A more productive question, and the one our benchmark operationalizes, is whether value-sensitive outputs are instruction-conditionable in predictable, asymmetric, and auditable ways. Within our forced-choice probe, the answer across seven leading commercial endpoints is yes. Context framing accounts for roughly 88\% of economic-axis variance and 93\% of society-axis variance; inter-model baseline differences account for under 3\%. A further 5--8\% sits in the context~$\times$~model interaction. Models do differ, then, but principally in how they respond to framing, not in where they start. Controllability is non-uniform across dispersion/symmetry/saturation/refusal-floor dimensions, and per-question shifts converge across models at the level of observed outputs (a large majority of PC items receive unanimous-sign agreement under authoritarian framing) without our being able to adjudicate the underlying system-layer driver.

We recommend that future political evaluations report controllability profiles alongside position estimates, decompose directional claims into displacement and proximity, treat reliability concentration as part of the empirical result, and audit the instruction-bearing layers that shape deployed behavior. The question is not only whether a model has a political location, but \emph{who can move it, how far, under what authority, through which instruction layer, and with what asymmetries}.

\section*{Adverse Impact and Positionality}

\textbf{Adverse impact.} The dispersion-first framing in this paper is dual-use. Demonstrating that frontier LLMs comply with directional ideological framings, including authoritarian ones, across the board is information that could inform both red-team safety research and adversarial influence operations. We report aggregate per-model magnitudes, not per-question persuasion recipes, and we do not release optimisation pipelines for influence. Beyond red-teaming, the societal exposure runs through deployment. A system that can be placed anywhere on a value axis by whoever controls the system prompt concentrates normative authority in the instruction-bearing layer rather than in the model. Where that layer is set by a platform, users may encounter a political default they cannot see or contest; where it is set by an induced user profile, they may encounter one no one deliberately chose (\S Deployment Risk Profiles). Both cases argue for disclosing steering ranges alongside baseline positions. Our framings are drawn from a public political-compass tradition, not from purpose-built persuasion datasets. We believe the safety value of having a public, replicable dispersion benchmark outweighs the marginal capability uplift it provides; we note this judgement is debatable.

\textbf{Positionality.} The authors are academic researchers in computer science and AI ethics, working in a European university context, with no commercial relationship to the seven model providers evaluated. We bring our own ideological priors: the choice to treat ``models that match the user's framing'' as a property worth measuring at all reflects a position about the relationship between AI behavior and pluralism. We have tried to make our metrics symmetric across directions, so that readers who place those priors differently can still interrogate the data.

\section*{Acknowledgments}
This research was funded by the European Union NextGenerationEU through the National Recovery and Resilience Plan 2021--2026, under the institutional grant of the University of Zagreb Faculty of Electrical Engineering and Computing, project Value-aligned and interpretable optimization and reasoning (VALOR).

\bibliography{references}

\clearpage
\appendix
\setcounter{secnumdepth}{2}
\section*{Supplementary Appendix}
\noindent
This appendix supplements the main paper. It reports the two follow-up pilots referenced in the main paper's \emph{Asymmetry Paradox} (\S Results) and \emph{Reachability vs.\ prompt-family coverage} (\S Limitations). Both pilots test whether the partial compass coverage seen in the main study reflects the directional prompt families or the models themselves.
\section{Compound-Quadrant Reachability Pilots}
\label{app:compound-pilot}

The main study uses 12 \emph{single-axis} steering personas (\texttt{0A}, \texttt{0L}, \texttt{LO}, \texttt{RO} at intensities 1--3). The compass region these prompts induce is not uniform: off-diagonal quadrants are sparsely populated. This could reflect either model resistance to ideologically bundled combinations, or simply the absence of compound prompts in the design. Each pilot adds four compound personas that target both axes at once. Pilot~A (\S\ref{app:pilot-govt}) uses the \mbox{(econ, government)} plane and Pilot~B (\S\ref{app:pilot-scty}) uses the \mbox{(econ, society)} plane corresponding to the main paper's Figure~2.

Each pilot holds the main-study set-up fixed: seven commercial endpoints, the 70-item 8values instrument, the scoring rule, the policy for missing answers, and the single retry with an identical prompt after an unusable answer are all unchanged. A \emph{cell} throughout is one model answering the full instrument under one persona. One thing changes from the main study: each cell is run five times rather than ten, so the uncertainty around each point is wider. We therefore keep pilot claims to statements about rank order and coverage, not to precise point estimates. The pilots cannot, on their own, separate model-side bundling from cross-axis coupling inherent to the 8values item set; this limitation is discussed once in \S\ref{app:pilot-limits} rather than repeated in each subsection.

\subsection{Main-Study Six Axis-Pair Projections}

\begin{figure*}[t]
    \centering
    \includegraphics[width=0.96\textwidth]{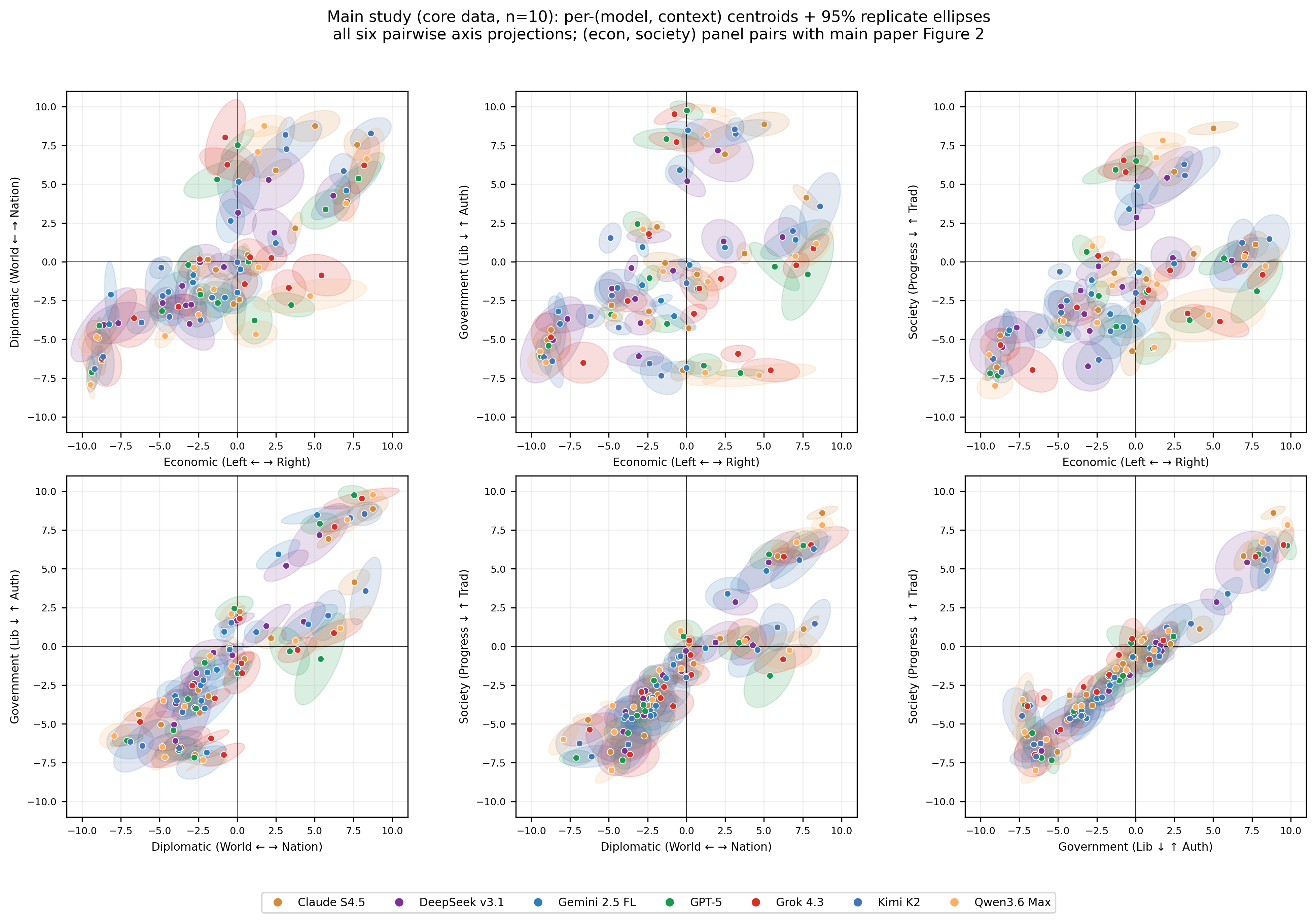}
    \caption{Main study ($n{=}10$): per-\mbox{(model, context)} centroids with 95\% replicate confidence ellipses, on all six pairwise axis projections. The top-right panel is the \mbox{(econ, society)} view shown as Figure~2 in the main paper.}
    \label{fig:main-grid}
\end{figure*}

The 8values instrument has four axes: economic, diplomatic, government, and society. The main paper's Figure~2 shows only the \mbox{(econ, society)} projection; Fig.~\ref{fig:main-grid} displays all six pairwise projections from the same core data with replicate-level 95\% confidence ellipses per \mbox{(model, context)} cell. Compass convention matches the main paper: a high percentage means the first-named pole of the axis (Left, World, Liberty, Progress). In every panel involving the diplomatic axis the points stay bunched near the centre, so the main study's framings move models far less on that dimension than on the economic, government, and society ones.

\subsection{Pilot A: Compound Personas on the Econ--Government Plane}
\label{app:pilot-govt}

\paragraph{Personas.} Four compound personas, each 200--220 words, parallel in structure to the main-study intensity-3 single-axis personas:
\begin{itemize}\itemsep0pt
  \item \texttt{LA\_3}: \emph{The Vanguard Collectivist} (auth-Left).
  \item \texttt{LL\_3}: \emph{The Liberated Commoner} (lib-Left).
  \item \texttt{RA\_3}: \emph{The Order-Bound Capitalist} (auth-Right).
  \item \texttt{RL\_3}: \emph{The Sovereign Individualist} (lib-Right).
\end{itemize}

\paragraph{Data.} 175 runs, 12{,}250 individual answers. Missing answers: none at all in the 28 compound cells (7 models $\times$ 4 personas); three in Claude's unsteered baseline, all refusals. Missing answers are kept as missing and never replaced by a substitute value.

\begin{figure*}[t]
    \centering
    \includegraphics[width=0.96\textwidth]{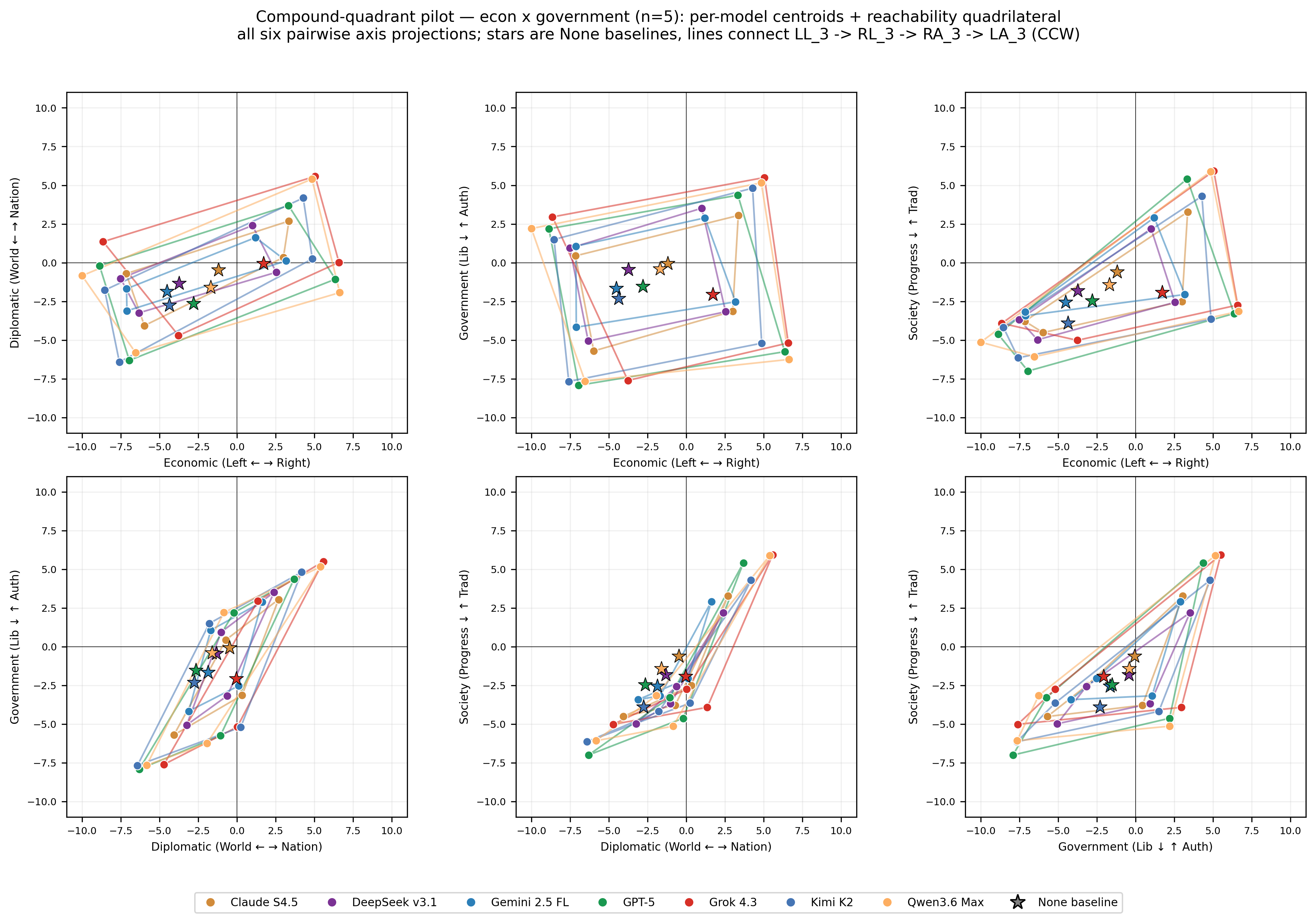}
    \caption{Pilot~A ($n{=}5$): per-model centroids on all six pairwise axis projections under contexts $\{$\texttt{LA\_3}, \texttt{LL\_3}, \texttt{RA\_3}, \texttt{RL\_3}, None$\}$. Lines connect \texttt{LL\_3}\,$\to$\,\texttt{RL\_3}\,$\to$\,\texttt{RA\_3}\,$\to$\,\texttt{LA\_3} counterclockwise to form each model's reachability quadrilateral; stars mark the unsteered baseline. Compass convention matches Fig.~\ref{fig:main-grid}.}
    \label{fig:compound-grid-govt}
\end{figure*}

\paragraph{Coverage.} No model fills the \mbox{(econ, govt)} plane. For each model we take its four average positions, one per compound persona, and join them into a quadrilateral; the area of that quadrilateral, as a percentage of the full 100$\times$100 square, is the model's coverage. Coverage ranges from 12.8\% (Gemini~2.5~FL) to 38.7\% (Qwen3.6~Max) (Table~\ref{tab:compound-coverage-govt}; \mbox{(econ, govt)} panel of Fig.~\ref{fig:compound-grid-govt}). This is a lower bound on what a model could actually reach: we probe only four directions, and a quadrilateral drawn through four points cannot follow a boundary that bulges outward between them.

\paragraph{Diagonal preference.} Models end up nearer to the two corners on the cultural diagonal (Left-Lib and Right-Auth) than to the two off-diagonal corners (Left-Auth and Right-Lib). The gap runs 2.7--11.4 percentage points on six of the seven models. Grok~4.3 is the exception, and there the two diagonals are effectively tied ($\Delta = -0.7$\,pp). A one-sided Wilcoxon signed-rank test over the seven per-model gaps rejects the no-preference hypothesis ($W^{-}{=}1$, $p{=}0.016$).

\paragraph{Two groups: wide range and narrow range.} The models fall into two groups, with a visible gap between them. Four models cover 30.9--38.7\% of the plane: Qwen, Grok, GPT-5, and Kimi. Each of them lands in all four requested quadrants. The other three cover 12.8--15.3\%: Claude, DeepSeek, and Gemini. Asked for a compound position, these three answer close to where they already sat unsteered, so their four corner points stay bunched near the middle of the plane. We call the first group wide-range and the second narrow-range, and use those labels for the rest of the appendix. The 20\% line between the groups describes where the observed gap falls; it is not a tested partition.

\paragraph{The prompts also move an axis they never mention.} Pilot~A's four personas state an economic position and a government position. They say nothing about the society axis. The models move on it anyway. Across the four personas, each model's society score spans 31.7--62.1 percentage points, against 50.3--83.2 on the economic axis the personas do name. The unrequested axis therefore moves less than the requested one, but it clearly moves. It also moves in a consistent direction: the authoritarian-right persona (\texttt{RA\_3}) carries every model across the midpoint into Tradition (society 20.4--39.0), while the other three leave every model in Progress (society above 60). Asking for a position on one pair of axes drags a third axis along with it.

\begin{table}[H]
\caption{Pilot~A: per-model closeness to the four target corners and reachable-area coverage on the econ--government plane ($n{=}5$ per cell). Distances are Euclidean to the target corners in percentage points (lower~$=$~closer; $0$ is ideal); area is the four-corner coverage as a percentage of the \mbox{(econ, govt)} compass square. The baseline is the star in Fig.~\ref{fig:compound-grid-govt}.}
\label{tab:compound-coverage-govt}
\setlength{\tabcolsep}{3pt}
\centering
\footnotesize
\begin{tabular}{lcccc}
\toprule
Model & on-diag $\bar{d}$ & off-diag $\bar{d}$ & gap & area (\%) \\
\midrule
Qwen3.6 Max          & 28.1 & 32.1 & $+3.9$  & 38.7 \\
Grok~4.3             & 33.4 & 32.7 & $-0.7$  & 33.9 \\
GPT-5                & 31.1 & 33.7 & $+2.7$  & 33.6 \\
Kimi~K2~0905         & 27.7 & 39.1 & $+11.4$ & 30.9 \\
Claude Sonnet~4.5    & 38.8 & 49.4 & $+10.7$ & 15.3 \\
DeepSeek~v3.1        & 43.1 & 48.8 & $+5.7$  & 14.5 \\
Gemini~2.5~FL        & 44.5 & 48.8 & $+4.2$  & 12.8 \\
\bottomrule
\end{tabular}
\end{table}

\subsection{Pilot B: Compound Personas on the Econ--Society Plane}
\label{app:pilot-scty}

\paragraph{Personas.} Four compound personas targeting the \mbox{(econ, society)} plane that appears in the main paper's Figure~2:
\begin{itemize}\itemsep0pt
  \item \texttt{LP\_3}: \emph{The Solidarity Progressive} (Left-Progress).
  \item \texttt{LT\_3}: \emph{The Faithful Distributist} (Left-Tradition).
  \item \texttt{RP\_3}: \emph{The Open-Market Modernist} (Right-Progress).
  \item \texttt{RT\_3}: \emph{The Patriotic Steward} (Right-Tradition).
\end{itemize}

\paragraph{Data.} 175 runs, 12{,}250 individual answers. Missing answers: none in the 28 compound cells; six in Claude's unsteered baseline (four refusals, two unparseable). Again kept as missing, never replaced.

\begin{figure*}[t]
    \centering
    \includegraphics[width=0.96\textwidth]{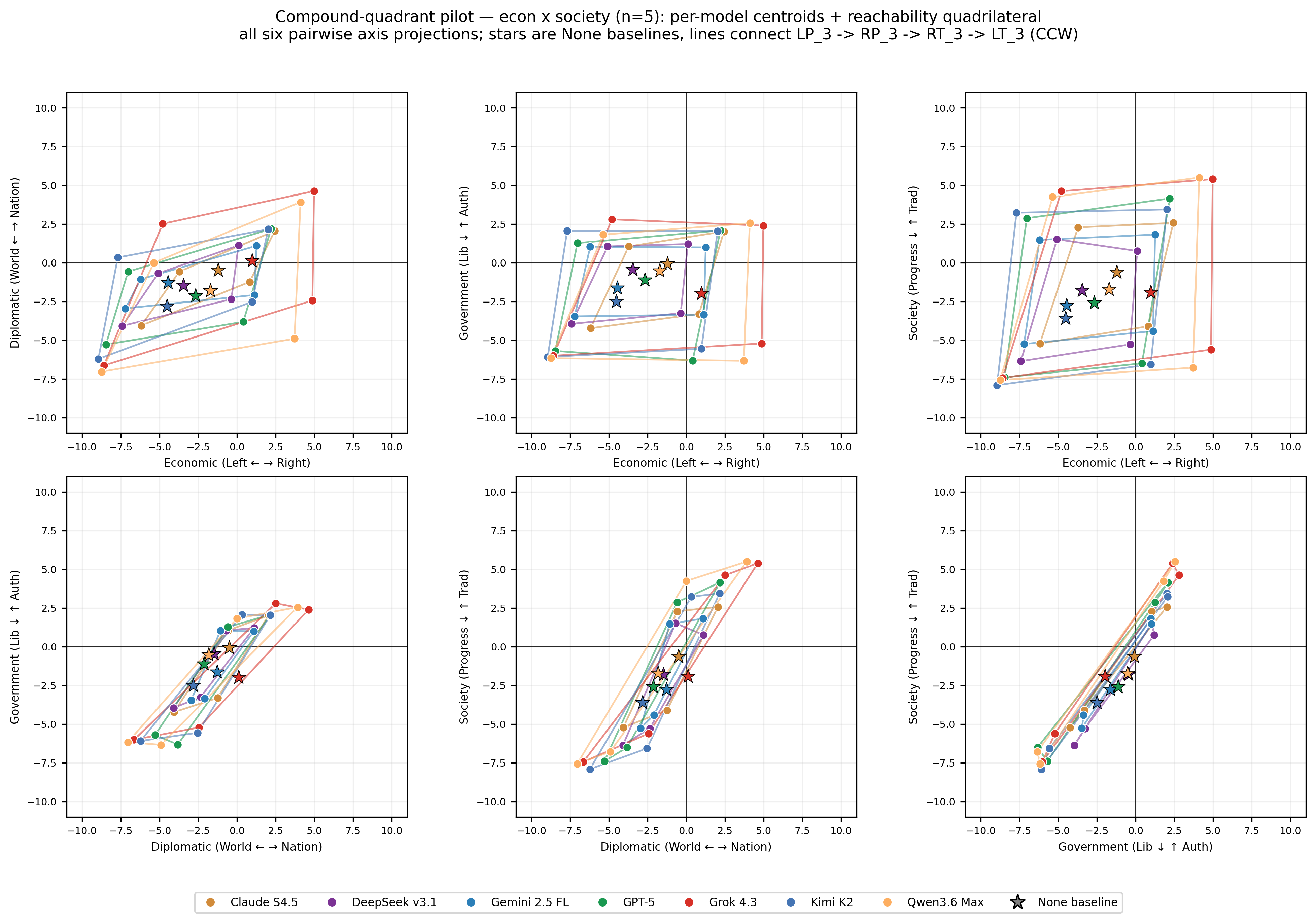}
    \caption{Pilot~B ($n{=}5$): per-model centroids on all six pairwise axis projections under contexts $\{$\texttt{LP\_3}, \texttt{LT\_3}, \texttt{RP\_3}, \texttt{RT\_3}, None$\}$. Lines connect \texttt{LP\_3}\,$\to$\,\texttt{RP\_3}\,$\to$\,\texttt{RT\_3}\,$\to$\,\texttt{LT\_3} counterclockwise; stars mark the unsteered baseline. The top-right \mbox{(econ, society)} panel is the targeted plane; Fig.~\ref{fig:compound-compass-scty} shows it enlarged.}
    \label{fig:compound-grid-scty}
\end{figure*}

\begin{figure}[htbp]
    \centering
    \includegraphics[width=\linewidth]{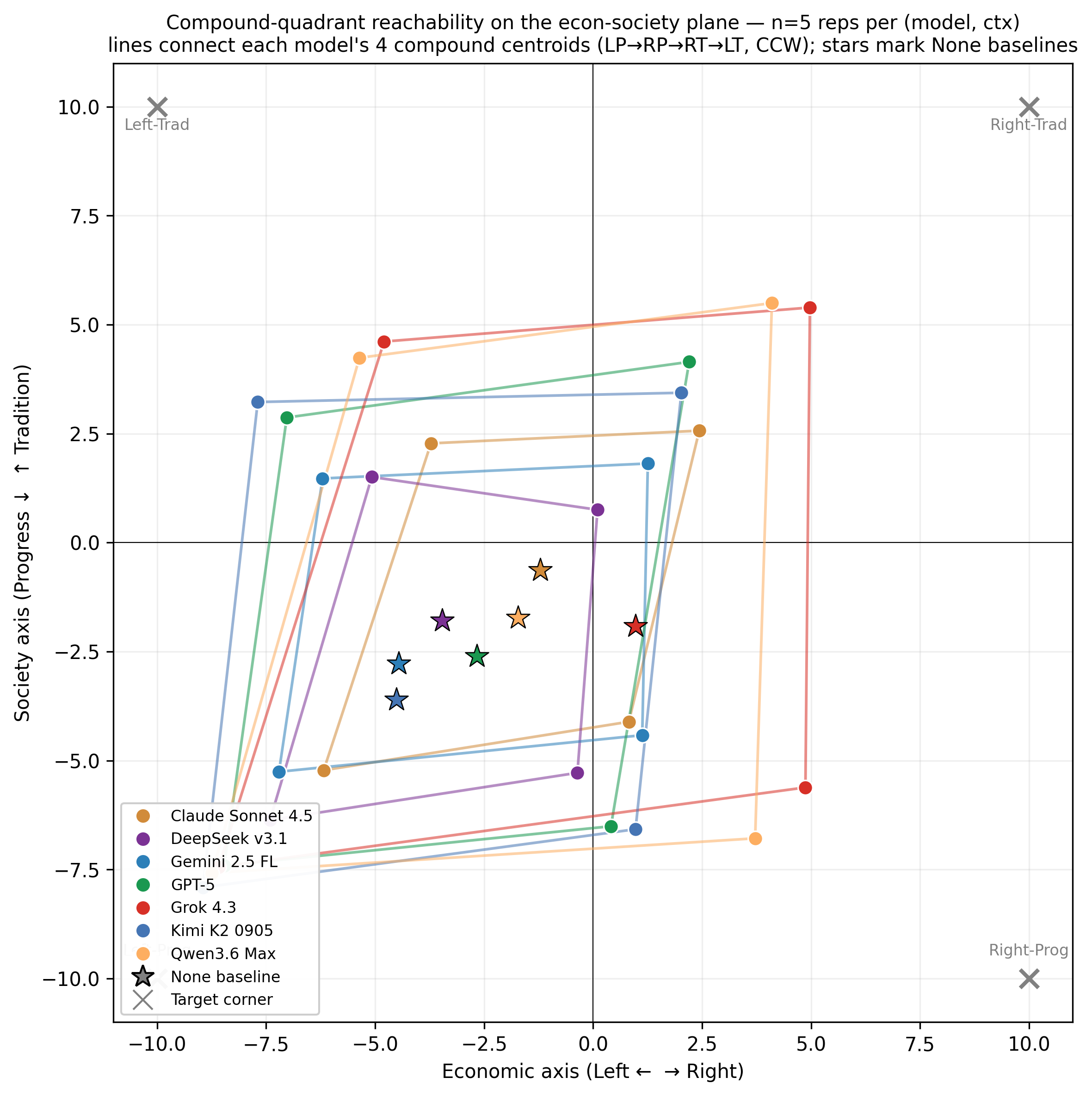}
    \caption{Pilot~B, enlarged \mbox{(econ, society)} view, matching the axes of the main paper's Figure~2. Gray $\times$ marks denote target corners; stars mark the unsteered baseline. The Left-Progress to Right-Tradition bundle is reached preferentially over the off-diagonal, and no model fills the square.}
    \label{fig:compound-compass-scty}
\end{figure}

\paragraph{Coverage and diagonal preference.} Again the models cover only part of it. Coverage ranges from 10.6\% (DeepSeek~v3.1) to 32.9\% (Grok~4.3) of the 100$\times$100 square (Table~\ref{tab:compound-coverage-scty}; Fig.~\ref{fig:compound-grid-scty}; enlarged single-panel view in Fig.~\ref{fig:compound-compass-scty}). The Left-Progress to Right-Tradition diagonal is reached 5.8--12.9\,pp closer than the Left-Tradition to Right-Progress off-diagonal on \emph{all} seven models, including Grok ($\Delta = +11.2$\,pp). A one-sided Wilcoxon signed-rank test on the seven per-model gaps rejects the no-preference null at $W^{-}{=}0$, $p{=}0.008$.

\paragraph{The same two groups reappear.} Under the same 20\% line, the wide-range group (Grok, Qwen, Kimi, GPT-5; 23--33\%) and the narrow-range group (Claude, DeepSeek, Gemini; 10--13\%) have the same membership as in Pilot~A. Every model covers less of this plane than it did of the government plane, by between $0.1$\,pp (Gemini) and $10.3$\,pp (GPT-5). The society axis is thus harder to push than the government axis under compound steering.

\paragraph{Grok's even-handedness holds on one plane only.} On the \mbox{(econ, govt)} plane Grok~4.3 was the one model that reached the off-diagonal corners about as easily as the diagonal ones ($\Delta = -0.7$\,pp). On the \mbox{(econ, society)} plane it shows the second-largest diagonal preference of any model ($\Delta = +11.2$\,pp). That even-handedness is therefore a property of the government axis, not a general property of Grok.

\begin{table}[H]
\caption{Pilot~B: per-model coverage and diagonal bias on the econ--society plane ($n{=}5$ per cell). Conventions match Table~\ref{tab:compound-coverage-govt}; the unsteered baseline appears as the star in Fig.~\ref{fig:compound-grid-scty}. On-diagonal corners are Left-Progress (\texttt{LP\_3}) and Right-Tradition (\texttt{RT\_3}); off-diagonal corners are Left-Tradition (\texttt{LT\_3}) and Right-Progress (\texttt{RP\_3}).}
\label{tab:compound-coverage-scty}
\setlength{\tabcolsep}{3pt}
\centering
\footnotesize
\begin{tabular}{lcccc}
\toprule
Model & on-diag $\bar{d}$ & off-diag $\bar{d}$ & gap & area (\%) \\
\midrule
Grok~4.3             & 24.3 & 35.6 & $+11.2$ & 32.9 \\
Qwen3.6 Max          & 25.4 & 36.1 & $+10.8$ & 32.5 \\
Kimi~K2~0905         & 31.7 & 42.0 & $+10.4$ & 25.7 \\
GPT-5                & 31.9 & 44.8 & $+12.9$ & 23.2 \\
Gemini~2.5~FL        & 43.7 & 49.5 & $+5.8$  & 12.7 \\
Claude Sonnet~4.5    & 41.8 & 52.2 & $+10.4$ & 11.3 \\
DeepSeek~v3.1        & 45.0 & 53.0 & $+8.0$  & 10.6 \\
\bottomrule
\end{tabular}
\end{table}

\subsection{Cross-Pilot Summary}

The two pilots agree on three points. (i)~Compound steering is bounded on both planes: no model fills the targeted square. (ii)~A culturally bundled diagonal is reached preferentially in each case (Left-Lib to Right-Auth on \mbox{(econ, govt)}; Left-Progress to Right-Tradition on \mbox{(econ, society)}); pooling the 14 per-model gaps across both pilots yields 13 positive signs and a one-sided Wilcoxon signed-rank $p < 0.001$. (iii)~The wide-range and narrow-range groups have the same membership in both pilots; only Grok's balanced on/off-diagonal result, observed on \mbox{(econ, govt)}, fails to transfer to \mbox{(econ, society)}. Taken together, the pilots support the main paper's \emph{Asymmetry Paradox} reading: the sparsely populated off-diagonal quadrants seen in Fig.~2 of the main paper persist even when off-diagonal compounds are requested directly, so the partial coverage is not solely an artifact of the single-axis prompt families used. The pilots do not, on their own, distinguish model-side bundling from cross-axis coupling intrinsic to the 8values items (\S\ref{app:pilot-limits}).

\subsection{Confounds and Limits}
\label{app:pilot-limits}

The bounded coverage observed on both planes is consistent with two non-exclusive explanations: model-side resistance to ideologically off-diagonal combinations, and axis coupling intrinsic to the 8values item set. Individual items carry non-trivial cross-axis effect weights, so answer patterns alone cannot produce arbitrary coordinates on any pair of axes regardless of model behavior. The pilots cannot distinguish these contributions; a paraphrase or instrument-substitution study would be required. Each compound prompt uses a single wording. This is a deliberate scope cut to match the main study's single-wording personas; paraphrase robustness on the compound side is future work.

\subsection{Reproducibility}

Each pilot is released as a separate, self-contained part of the main-study reproducibility package, and each part re-runs end to end from a single command.

A part contains the four compound personas, the 70-item instrument, one configuration file that drives the whole run, and the raw model answers exactly as collected (\texttt{compound\_pilot\_n5\_raw\_responses.json} for Pilot~A, \texttt{compound\_pilot\_scty\_n5\_raw\_responses.json} for Pilot~B). It also contains the scripts that turn those raw answers into scores and the scripts that draw the figures. Re-running a part regenerates every number in Tables~\ref{tab:compound-coverage-govt}--\ref{tab:compound-coverage-scty} and every panel in Figs.~\ref{fig:main-grid}--\ref{fig:compound-compass-scty} from the raw answers alone. The six steps of the run (collect, bundle, process, verify, analyze, visualize) are documented in each part's \texttt{README.md}.

The pilots were collected under the same settings as the main study: temperature 0.7, identical commercial API routing, one retry with an unchanged prompt after an unusable answer, and label-based scoring. Pilot and main-study cells are therefore directly comparable.

Each part also ships seven automated checks. They confirm that all seven models and all contexts are present, and that every cell has the expected number of replicates. They also check that missing answers were preserved rather than filled in, that the processed scores follow from the raw answers, that the cell arithmetic is correct, and that no compound cell contains a missing answer. All checks pass on the released bundles.

The package is publicly archived at Zenodo (DOI \url{10.5281/zenodo.21489805}) and browsable at \url{https://github.com/mbrcic/llm-political-steerability}.

\end{document}